\documentclass[aps,prl,twocolumn,superscriptaddress,showpacs,showkeys]{revtex4}

\usepackage{graphicx}
\usepackage{bm}
\usepackage{amsmath}

\begin{document}

\title{Bose-Einstein condensate coupled to a nanomechanical resonator on an atom chip}

\author{Philipp~Treutlein}\email[Electronic address: ]{treutlein@lmu.de}
\author{David~Hunger}
\author{Stephan~Camerer}
\author{Theodor~W.~H{\"a}nsch}
\affiliation{Max-Planck-Institut f{\"u}r Quantenoptik and Fakult{\"a}t f{\"u}r Physik der Ludwig-Maximilians-Universit{\"a}t, Schellingstr.~4, 80799 M{\"u}nchen, Germany}
\author{Jakob~Reichel}
\affiliation{Laboratoire Kastler Brossel de l'E.N.S., 24 Rue Lhomond, 75231 Paris Cedex 05, France}

\date{\today}

\begin{abstract}
We theoretically study the coupling of Bose-Einstein condensed atoms to the mechanical oscillations of a nanoscale cantilever with a magnetic tip. This is an experimentally viable hybrid quantum system which allows one to explore the interface of quantum optics and condensed matter physics. We propose an experiment where easily detectable atomic spin-flips are induced by the cantilever motion. This can be used to probe thermal oscillations of the cantilever with the atoms. At low cantilever temperatures, as realized in recent experiments, the backaction of the atoms onto the cantilever is significant and the system represents a mechanical analog of cavity quantum electrodynamics. With high but realistic cantilever quality factors, the strong coupling regime can be reached, either with single atoms or collectively with Bose-Einstein condensates. We discuss an implementation on an atom chip.
\end{abstract}


\maketitle

Quantum optics and condensed matter physics show a strong convergence.
On the one hand, quantum optical systems, most notably neutral atoms in optical lattices, have been used to experimentally investigate concepts of condensed matter physics such as Bloch oscillations and Fermi surfaces \cite{Bloch05}.
On the other hand, micro- and nanostructured condensed matter systems enter a regime described by concepts of quantum optics, as exemplified by circuit cavity quantum electrodynamics \cite{Wallraff04}, laser-cooling of mechanical resonators \cite{CoolNano06}, and measurement backaction effects in cryogenic mechanical resonators \cite{Naik06}.
A new exciting possibility beyond this successful conceptual interaction is to {\em physically couple} a quantum optical system to a condensed matter system. Such a hybrid quantum system can be used to study fundamental questions of decoherence at the transition between quantum and classical physics and has possible applications in precision measurement \cite{Wang06} and quantum information processing \cite{Tian04}.

Atom chips \cite{Fortagh07} are ideally suited for the implementation of hybrid quantum systems. Neutral atoms can be positioned with nanometer precision \cite{Hommelhoff05} and trapped at distances below $1~\mu$m from the chip surface \cite{Lin04}. Coherent control of internal \cite{Treutlein04} and motional \cite{Hofferberth06} states of atoms in chip traps is a reality.
Atom-surface interactions are sufficiently understood \cite{Fortagh07} so that undesired effects can be mitigated by choice of materials and fabrication techniques.
This is an advantage over systems such as ions or polar molecules on a chip, which have recently been considered in this context \cite{Tian04, Andre06}.
A first milestone is to realize a controlled interaction between atoms and a nanodevice on the chip surface.

In this paper, we investigate magnetic coupling between the spin of atoms in a Bose-Einstein condensate (BEC) \cite{Haensel01a} and a single vibrational mode of a nanomechanical resonator \cite{Ekinci05} on an atom chip. We find that the BEC can be used as a sensitive quantum probe which allows one to detect the thermal motion of the resonator at room temperature.
At lower resonator temperatures, the backaction of the atoms onto the resonator is significant and the coupled system realizes a mechanical analog of cavity quantum electrodynamics (cQED) in the strong coupling regime.
We specify in detail a realistic setup for the experiment, which can be performed with available atom chip technology, and thus allows one to explore this fascinating field already today.

\begin{figure}[b]
\includegraphics[width=0.48\textwidth]{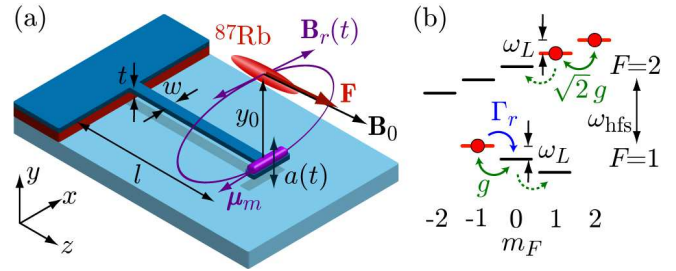}
\caption{\label{fig:Setup} (color online) BEC-resonator coupling mechanism. (a) Atom chip with a BEC of $^{87}$Rb atoms (red: BEC wave function) at a distance $y_0$ from a nanomechanical resonator. The free-standing structure (dark blue) is supported at one end to form a cantilever-type resonator that performs out-of-plane mechanical oscillations $a(t)$. The single-domain ferromagnet (purple) on the resonator tip creates a magnetic field with oscillatory component $\mathbf{B}_r(t)$
which couples
to the atomic spin $\mathbf{ F}$. (b) Hyperfine structure of $^{87}$Rb in the magnetic field $\mathbf{B}_0$. Hyperfine levels $|F,m_F\rangle$ are coupled (blue or green arrows, depending on experiment) if the Larmor frequency $\omega_L$ is tuned to the oscillation frequency of the resonator.
Magnetically trappable states indicated in red.}
\end{figure}

The physical situation is illustrated in Fig.~\ref{fig:Setup}(a).
$^{87}$Rb atoms are trapped in a magnetic microtrap at a distance $y_0$ above a cantilever resonator, which is nanofabricated on the atom chip surface. The cantilever tip carries a single-domain ferromagnet which creates a magnetic field 
with a strong gradient $G_m$.
The magnet transduces out-of-plane mechanical oscillations $a(t)=a\cos(\omega_r t + \varphi)$ of the cantilever tip of amplitude $a \ll y_0$ into an oscillatory magnetic field
$\mathbf{B}_r(t)=G_m a(t) \mathbf{e}_x$ in the center of the microtrap. The orientation of the magnet is chosen such that $\mathbf{B}_r$ is perpendicular to the static magnetic field $\mathbf{B}_0=B_0 \mathbf{e}_z$ in the trap center. The atomic spin $\mathbf{ F}$ interacts with $\mathbf{B}_r(t)$ via the Zeeman Hamiltonian
\begin{equation}\label{eq:HZ}
 H_Z = - \boldsymbol{ \mu} \cdot \mathbf{B}_r(t) = \mu_B g_F  F_x G_m a(t),
\end{equation}
where $\boldsymbol{ \mu} = -\mu_B g_F \mathbf{ F}$ is the operator of the magnetic moment. In this way, the ferromagnet establishes a coupling between the spin and the resonator mechanical motion.

The ground state hyperfine spin levels $|F,m_F\rangle$ of $^{87}$Rb are shown in Fig.~\ref{fig:Setup}(b). The energy splitting between adjacent $m_F$-levels is given by the Larmor frequency
$\omega_L = \mu_B |g_F| B_0 / \hbar$.
Note that $\omega_L$ is widely tunable by adjusting $B_0$.
This allows one to control the detuning $\delta=\omega_r-\omega_L$ between a given resonator mode of frequency $\omega_r$ and the atomic resonance in the trap center. Quickly changing $\delta$ switches the coupling on and off. Near resonance ($\delta \approx 0$), the coupling leads to spin flips.

In a magnetic trap, only weak-field seeking states are trapped, as indicated in Fig.~\ref{fig:Setup}(b). This can be exploited in a simple way to detect the spin flips induced by the coupling: atoms initially trapped in state $|1,-1\rangle$ are coupled by the nanoresonator to $|1,0\rangle$, where they are quickly lost from the trap. This is analogous to a cw atom laser experiment \cite{Bloch99}, with the mechanical resonator inducing the radio-frequency magnetic field for output coupling. The rate $\Gamma_r$ at which atoms are coupled out of the BEC is a sensitive probe revealing the temporal dynamics of the resonator motion.

To derive $\Gamma_r$, we follow the theory of \cite{Steck98}, which includes effects of interatomic interactions, but neglects gravity. This is justified here due to the high trap frequencies. The trapped BEC in $|1,-1\rangle$ is assumed to be in the Thomas-Fermi (TF) regime. It is coupled with Rabi frequency $\Omega_R = \mu_B G_m a/\sqrt{8}\hbar $ to $|1,0\rangle$, where a continuum of untrapped motional states is available to the atoms. The energy width of this continuum is given by the BEC chemical potential $\mu_c$. For typical parameters (see below), $\hbar \Omega_R \ll \mu_c$, and only a fraction $\simeq \hbar \Omega_R/\mu_c$ of the BEC atoms is resonantly coupled. In this limit \cite{Steck98},
\begin{equation}\label{eq:Gammar}
\Gamma_r = \frac{15 \pi}{8} \frac{\hbar \Omega_R^2}{\mu_c} \left(r_c-r_c^3\right),
\end{equation}
where $r_c = \sqrt{\hbar \delta / \mu_c}$. Output coupling takes place on a thin ellipsoidal shell of resonance with main axes $r_i = r_c R_i$, where $R_i$ are the TF radii of the BEC.


\begin{figure}[t]
    \centering
        \includegraphics[width=0.48\textwidth]{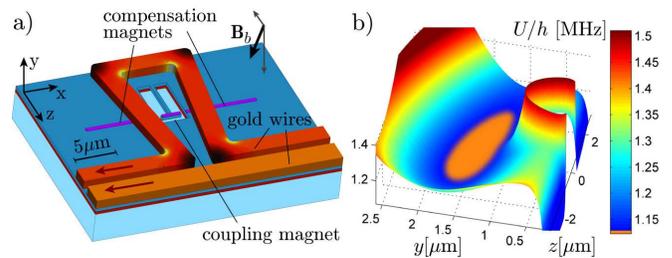}
    \caption{\label{fig:Chiplayout} (color online) (a) Atom chip layout. The trapping potential is created by gold wires (width $2\,\mu$m, current $4.4\,$mA) and a homogeneous field $\mathbf{B}_b=(-0.1,-4.2,-1.6)\,$G. The wire color indicates the current density obtained from a finite elements simulation. The Co coupling magnet on the tip of the Si cantilever is located directly below the center of the atom trap. Compensation magnets on each side of the tip reduce distortion of the trapping potential due to the static field of the coupling magnet. (b) Trapping potential in the $yz$-plane intersecting the resonator. Trap minimum at $(y_0,z_0)=(1.5,0.0)\,\mu$m, trap frequencies $\omega_{x,y,z}/2\pi=(8.9,9.7,1.2)\,$kHz for $|1,-1\rangle$. The static field of the magnets causes a repulsive potential around $y=z=0$. The attractive Casimir-Polder surface potential is visible for $y \rightarrow 0$. The orange area in the trap center shows the extension of the BEC.}
\end{figure}
Figure~\ref{fig:Chiplayout}(a) shows an implementation which we envisage in collaboration with the Kotthaus group at CeNS \cite{Scheible04}. The atom chip is fabricated by several steps of e-beam lithography on a Si-on-insulator wafer in combination with lift-off metallization and selective etching of the oxide to create free-standing Si structures.
Atoms are trapped in a Ioffe-type trap created by currents in the chip wires \cite{Fortagh07}.
Our numerical simulation of the trapping potential (Fig.~\ref{fig:Chiplayout}(b)) includes the magnetic fields of the wires, the Co magnets, the homogeneous field $\mathbf{B}_b$, gravity, and the Casimir-Polder surface potential \cite{Fortagh07}.
Trap frequencies are adjustable in the kHz-range, the aspect ratio is $\omega_z/\omega_{x,y}\approx 0.1$.
The background trap loss rate $\gamma = \gamma_\textrm{tbl} + \gamma_0$ is dominated by three-body collisions \cite{Soeding99} with $\gamma_\textrm{tbl} = 2.2 \times 10^{-12}\,\mathrm{s}^{7/5} \times \bar{\omega}_t^{12/5}N^{4/5}$ for $|1,-1\rangle$, where $\bar{\omega}_t = (\omega_x \omega_y \omega_z)^{1/3}$. Background gas collisions and atom-surface interactions contribute a much smaller rate $\gamma_0$.


The ferromagnet is a single magnetic domain whose magnetic moment $\boldsymbol{\mu}_m$ is spontaneously oriented along its long axis due to the shape anisotropy. For Co nanobars, a switching field $> 500$\,G \cite{Kong96} ensures that the magnetization of the bar is nearly unaffected by the fields applied for magnetic trapping, which are $< 100$\,G.
Approximating the bar by a magnetic dipole, we have $G_m = 3\mu_0 |\boldsymbol{\mu}_m| / 4 \pi y_0^4$. By changing $y_0$ and the magnet dimensions, $G_m$ can be adjusted. Equation~(\ref{eq:HZ}) suggests that the strength of the atom-nanoresonator coupling can be maximized by increasing $G_m$ as much as possible. However, the atoms experience a force in this field gradient, and an excessively large $G_m$ would strongly distort the trapping potential. To mitigate distortion, two compensation magnets are placed next to the coupling magnet with identical direction of magnetization. This reduces the static field gradient at the location of the atoms, while the oscillatory field $\mathbf{B}_r(t)$ remains unaffected as the compensation magnets do not oscillate.


Nanomechanical resonators have a complex spectrum of vibrational modes. Due to their high quality factors $Q = 10^3 - 10^5$ \cite{Ekinci05,Naik06}, the modes are well resolved.
The BEC is coupled to the fundamental out-of-plane flexural oscillation at frequency $\omega_r/2\pi \approx 0.16 \sqrt{E /\rho (1+c)}\,(t/l^2)$. Here, $t \leq w \ll l$ are the dimensions, $E$ is Young's modulus, and $\rho$ is the mass density of the cantilever, while $c=m/0.24 \rho l w t$ accounts for the additional mass $m$ of the magnet and the Si paddle at the cantilever tip.
The force acting on the coupling magnet in the magnetic field of the compensation magnets leads to an additional shift of $\omega_r$ which is determined numerically and included in the numbers given below \cite{eddy}.
We model the cantilever tip as a harmonic oscillator of frequency $\omega_r$ with an effective mass $m_\textrm{eff} \approx 0.24 \rho l w t + m$, obtained by integrating over the mode function.

\begin{figure}[t]
    \centering
        \includegraphics[width=0.45\textwidth]{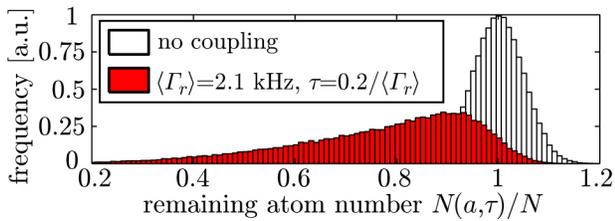}
    \caption{\label{fig:TrapLoss} (color online) Coupling of the BEC to a thermally driven cantilever at $T=300$\,K. Simulated histogram of the fraction of atoms remaining in the trap after a time $\tau$, including background loss. For comparison, the atom number distribution without coupling is shown. We have assumed 5\% fluctuations in atom number due to technical noise.}
\end{figure}

In contrast to the BEC, which is a prime example of quantum-mechanical coherence, dissipation and thermal effects play an important role in the cantilever dynamics.
In thermal equilibrium with its environment at temperature $T$, the cantilever performs oscillations at frequency $\omega_r$ with random amplitude $a$ and phase $\varphi$ \cite{Briant03}.
Both $a$ and $\varphi$
change on a timescale $\kappa^{-1}$, where $\kappa=\omega_r/2Q$ is the damping rate. For a high-$Q$ cantilever, however, this timescale is longer than the time scale $\Gamma_r^{-1}$ of coupling to the BEC, as we will show below. This allows one to use the BEC as a probe to directly monitor the thermal fluctuations. In a single shot of such an experiment, the cantilever performs simple harmonic motion with constant $a$ and $\varphi$. The BEC in $|1,-1\rangle$ is coupled to the cantilever for a time $\tau \ll \kappa^{-1}$ and the remaining number of atoms $N(a,\tau) = N \exp[-\Gamma_r(a)\, \tau]$ is measured. Repeating the experiment, one observes fluctuations of $N(a,\tau)$ due to the fluctuations of $a$.
Figure~\ref{fig:TrapLoss} shows a simulated histogram of $N(a,\tau)/N$. Since $\Gamma_r \propto a^2 \propto n$, the histogram reflects the exponential distribution of phonon numbers $n$ in the thermal state of the resonator, with $\langle \Gamma_r \rangle$ given by the mean phonon number $n_\textrm{th} = [\exp(\hbar \omega_r / k_B T)-1]^{-1}$.

As a realistic example, we take a Si cantilever with $(l,w,t) = (7.0, 0.2, 0.1)\,\mu$m, $m_\textrm{eff} = 3\times 10^{-16}$\,kg, and $\omega_r/2\pi = 1.12\,$MHz.
It carries a Co magnet of dimensions $(l_m,w_m,t_m)=(1.3,0.2,0.08)\,\mu$m; the two compensation magnets have the same cross section and $5\,\mu$m length, the gap between magnets is $d=200\,$nm. Trap parameters optimizing $\langle \Gamma_r \rangle/\gamma$ for a BEC of $N=10^3$ atoms are given in Fig.~(\ref{fig:Chiplayout}). For given $\bar\omega_t$, we adjust $G_m$ to the maximum value allowed by trap distortion \cite{Gm}.
The mean coupling rate for $r_c = 1/\sqrt{3}$ is $\langle \Gamma_r\rangle = 2.1$\,kHz. Background losses are much smaller, $\gamma = 0.01\,\langle\Gamma_r\rangle$. Taking a moderate $Q=5\times 10^3$ and $\tau = 0.2/\langle\Gamma_r\rangle$, we have $\kappa \tau = 0.07$. This shows that coupling the BEC to the thermal motion of the resonator is easily achieved with parameters which are based on fabrication results at CeNS.


At room temperature, the thermal equilibrium state of the resonator has an average phonon number $n_\textrm{th} \gg N$, and the coupling does not significantly perturb the state of the resonator.
By cooling a resonator in a dilution refrigerator, phonon numbers as low as $n_\textrm{th} = 25$ have been observed \cite{Naik06}. Recently, laser cooling of mechanical oscillations was demonstrated \cite{CoolNano06}, which opens the exciting perspective of preparing a single mode of the resonator with very low $n_\textrm{th}$ or even reaching the quantum mechanical ground state ($n_\textrm{th} \ll 1$) without a cryostat.
At low mode temperatures, $n_\textrm{th} \sim N$ and the backaction of the BEC onto the resonator cannot be neglected. Every atom changing its state changes the number of phonons in the resonator mode by one.
In this regime, it is possible to use the BEC as an actuator for the mechanical oscillations. The two systems exchange energy coherently, increasing or decreasing the number of phonons depending on the initial state of the BEC.

In analogy with cQED, we derive a fully quantum-mechanical theory for the dynamics of the coupled system.
We now consider a transition between two trapped atomic states $|0\rangle \leftrightarrow |1\rangle$. In a magnetic trap, $|0\rangle \equiv |2,1\rangle$ and $|1\rangle \equiv |2,2\rangle$ can be used (see Fig.~\ref{fig:Setup}(b)). However, the different trap frequencies lead to entanglement between internal and motional atomic degrees of freedom. Here, we discuss the simpler situation of an optical or electrodynamic microtrap \cite{Kishimoto06}, which provides identical trapping potentials for all hyperfine states. In such a trap, all atoms in the BEC couple simultaneously to the resonator. Since collisional losses are lower in $F=1$, we choose $|0\rangle \equiv |1,0\rangle$ and $|1\rangle \equiv |1,-1\rangle$.
The transition $|0\rangle \leftrightarrow |1\rangle$ can be decoupled from other $m_F$-levels by making use of the quadratic Zeeman effect or by using microwaves to induce $m_F$-dependent energy shifts \cite{Treutlein06b}.
A BEC of $N$ two-level atoms with level spacing $\hbar \omega_L$ can be described by a collective spin $S=N/2$ with Hamiltonian $ H_\textrm{BEC}=\hbar \omega_L  S_z$ and eigenstates $|S,m_S\rangle,\,|m_S|\leq S$ \cite{Shore93}.
The Hamiltonian of the quantized resonator is $ H_r = \hbar \omega_r  a^+  a$, where $ a$ ($ a^+$) is the annihilation (creation) operator for phonons in the fundamental mechanical mode.
The coupling Hamiltonian is obtained by replacing $\sqrt{2}\, g_F  F_x \rightarrow  S_x$ and $a(t) \rightarrow a_\textrm{qm}( a^+ +  a)$ in Eq.~(\ref{eq:HZ}), where
$a_\textrm{qm} = \sqrt{\hbar/2 m_\textrm{eff} \omega_r}$
is the r.m.s. amplitude of the quantum mechanical zero-point motion.
For the coupled system, $H=H_r +  H_\textrm{BEC} +  H_Z$. With $ S^\pm =  S_x \pm i  S_y$ and applying the rotating-wave approximation, we obtain the Tavis-Cummings Hamiltonian \cite{Shore93},
\begin{equation}\label{eq:HJC}
  H = \hbar \omega_r  a^+  a  + \hbar \omega_L  S_z + \hbar g ( S^+  a  +  S^- a^+),
\end{equation}
where
$g=\mu_B G_m a_\textrm{qm}/\sqrt{8} \hbar$
is the single-atom single-phonon coupling constant.
In cQED, Eq.~(\ref{eq:HJC}) usually describes the coupling of atoms to the electromagnetic field of a single mode of an optical or microwave cavity \cite{Shore93}. Here, it describes the coupling to the phonon field of a mode of mechanical oscillations. In this sense, this is a mechanical cQED system.

We now investigate whether the strong coupling regime of cQED \cite{Kimble94} can be reached, where coherent dynamics occurs at a faster rate than dissipative dynamics.
For a single atom ($N=1$), this requires $g>(\kappa,\gamma)$.
We maximize $g/(\kappa+\gamma)$ by optimizing resonator, magnet, and trap parameters.
We take a Si cantilever with $(l,w,t) = (8.0, 0.3, 0.05)\,\mu$m and assume $Q=10^5$ as in recent experiments at low $T$ \cite{Naik06}. It carries a Co magnet with $(l_m,w_m,t_m)=(250,50,80)\,$nm and $d=40\,$nm, resulting in $\omega_r/2\pi = 2.8\,$MHz. For $N=1$ there is not collisional loss; therefore, higher $\bar\omega_t$ is possible.
In a trap with $\bar\omega_t/2\pi=250$\,kHz and $y_0=250$\,nm, realistic on atom chips \cite{Katori04}, we obtain strong coupling with $(g,\kappa,\gamma)=2\pi \times (62, 14, 0.3)\,$Hz.
A related quantity is the cooperativity parameter, $C = g^2/ 2\kappa \gamma$. For $C > 1$, mechanical analogs of optical bistability and squeezing can be observed. We obtain $C = 430$.
To prepare the resonator with $n_\textrm{th}<1$, $T<0.1$\,mK is required. Such low temperatures could perhaps be achieved by laser cooling.

For $N$ atoms identically coupled to the resonator, the coupling is collectively enhanced. Coherent dynamics now occurs if the weaker condition $g\sqrt{N} > (\kappa, \gamma)$ is met \cite{Kimble94}. This is still true for a resonator in a thermal state as long as $n_\textrm{th} \ll N$, putting less stringent limits on $T$.
For resonator dimensions as above, $(l_m,w_m,t_m)=(2.0,0.06,0.12)\,\mu$m, and $d=100\,$nm we obtain $\omega_r/2\pi = 1.1\,$MHz. Maximizing $g\sqrt{N}/(\kappa+\gamma)$ for a BEC with $N=10^4$ atoms, we find a trap with $\bar{\omega}_t/2\pi = 2.9$\,kHz and $y_0=2.0\,\mu$m.
At $T=50$\,mK (typical in a dilution refrigerator), $n_\textrm{th} \approx 980 \ll N$. Collective strong coupling is reached with $(g\sqrt{N},\kappa,\gamma) = 2\pi\times (21,5,10)$\,Hz and the $N$-atom cooperativity is $C N = 4$.

In a quantum Monte Carlo simulation, we couple a BEC in state $|S=N/2,m_S=N/2\rangle$ (i.e.\ all atoms in state $|1\rangle$) to a resonator with $n_\textrm{th} \ll N$. The coupling drives the resonator out of thermal equilibrium into a state with a mean phonon number $\langle n \rangle \gg n_\textrm{th}$. Conversely, if the BEC is prepared in state $|S=N/2,m_S=-N/2\rangle$ (all atoms in state $|0\rangle$), excitations are initially transferred from the resonator to the BEC, creating a state with $\langle n \rangle \ll n_\textrm{th}$. The time scale for both processes is $\pi/2 g \sqrt{N}$. Depending on the initial conditions, the BEC can therefore be used to drive or cool the resonator mode.

%

We have shown that a BEC on an atom chip can be used as a sensitive probe, as a coolant, and as a coherent actuator for a nanomechanical resonator. The coupling could be used to transfer nonclassical states of the BEC to the mechanical system. Due to the dissipative coupling of the resonator to its environment, interesting questions of decoherence arise and can be studied with this system.
Instead of coupling different spin levels, it is also possible to couple the resonator to the motional degrees of freedom of either a BEC or a single atom, similar to the coupling mechanism proposed for a nanoscale ion trap in \cite{Tian04}.
In a recent experiment, a spin resonance transition in a thermal atomic vapor was excited by a driven mechanical resonator with a magnetic tip \cite{Wang06}. We expect that the system considered here, though requiring a smaller resonator and much better control over the atoms, can be realized in the near future.

We thank our collaborators D.~K{\"o}nig, D.~Scheible, F.~Beil, and J.~Kotthaus at CeNS, and F.~Marquardt, P.~Rabl, and P.~Zoller for stimulating discussions. Research supported by the Nanosystems Initiative Munich.


\begin{thebibliography}{29}

\bibitem{Bloch05}
I.~Bloch, Nat. Phys. {\bfseries 1}, 23 (2005).

\bibitem{Wallraff04}
A.~Wallraff {\itshape et al.}, Nature {\bfseries 431}, 162 (2004).

\bibitem{CoolNano06}
C.~H{\"o}hberger Metzger and K.~Karrai, Nature {\bfseries 432}, 1002 (2004);
O.~Arcizet {\itshape et al.}, {\itshape ibid.} {\bfseries 444}, 71 (2006);
S.~Gigan {\itshape et al.}, {\itshape ibid.} {\bfseries 444}, 67 (2006);
D.~Kleckner and D.~Bouwmeester, {\itshape ibid.} {\bfseries 444}, 75 (2006);
A.~Schliesser {\itshape et al.}, Phys. Rev. Lett. {\bfseries 97}, 243905 (2006);
I.~Favero {\itshape et al.}, Appl. Phys. Lett. {\bfseries 90}, 104101 (2007).

\bibitem{Naik06}
A.~Naik {\itshape et al.}, Nature {\bfseries 443}, 193 (2006).

\bibitem{Wang06}
Y.-J.~Wang {\itshape et al.}, Phys. Rev. Lett. {\bfseries 97}, 227602 (2006).

\bibitem{Tian04}
L.~Tian and P.~Zoller, Phys. Rev. Lett. {\bfseries 93}, 266403 (2004).

\bibitem{Fortagh07}
J.~Fort{\'a}gh and C.~Zimmermann, Rev. Mod. Phys. {\bfseries 79}, 235 (2007).

\bibitem{Hommelhoff05}
P.~Hommelhoff {\itshape et al.}, New J. Phys. {\bfseries 7}, 3 (2005).

\bibitem{Lin04}
Y.~Lin, I.~Teper, C.~Chin, and V.~Vuleti{\'c}, Phys. Rev. Lett. {\bfseries 92}, 050404 (2004).

\bibitem{Treutlein04}
P.~Treutlein {\itshape et al.}, Phys. Rev. Lett. {\bfseries 92}, 203005 (2004).

\bibitem{Hofferberth06}
S.~Hofferberth {\itshape et al.}, Nat. Phys. {\bfseries 2}, 710 (2006).


\bibitem{Andre06}
A.~Andr{\'e} {\itshape et al.}, Nat. Phys. {\bfseries 2}, 636 (2006).

\bibitem{Haensel01a}
W.~H{\"a}nsel, P.~Hommelhoff, T.~W.~H{\"a}nsch, and J.~Reichel, Nature {\bfseries 413}, 498 (2001).

\bibitem{Ekinci05}
K.~L.~Ekinci and M.~L.~Roukes, Rev. Sci. Instrum. {\bfseries 76}, 061101 (2005).

\bibitem{Bloch99}
I.~Bloch, T.~W.~H{\"a}nsch, and T.~Esslinger, Phys. Rev. Lett. {\bfseries 82}, 3008 (1999).

\bibitem{Steck98}
H.~Steck, M.~Naraschewski, and H.~Wallis, Phys. Rev. Lett. {\bfseries 80}, 1 (1998).

\bibitem{Scheible04}
D.~V.~Scheible, C.~Weiss, J.~P.~Kotthaus, and R.~H.~Blick, Phys. Rev. Lett. {\bfseries 93}, 186801 (2004).

\bibitem{Soeding99}
J.~S{\"o}ding {\itshape et al.}, Appl. Phys. B {\bfseries 69}, 257 (1999).

\bibitem{Kong96}
L.~Kong and S.~Y.~Chou, J. Appl. Phys. {\bfseries 80}, 5205 (1996).


\bibitem{eddy}
We have estimated that cantilever damping due to this interaction as well as due to eddy currents induced in the gold wires is negligible in our geometry.

\bibitem{Briant03}
T.~Briant, P.~F.~Cohadon, M.~Pinard, and A.~Heidmann, Eur. Phys. J. D {\bfseries 22}, 131 (2003).

\bibitem{Gm}
In calculating $G_m$, we use the formula of R.~Engel-Herbert and T.~Hesjedal, J. Appl. Phys. {\bfseries 97}, 074504 (2005).

\bibitem{Kishimoto06}
T.~Kishimoto {\itshape et al.}, Phys. Rev. Lett. {\bfseries 96}, 123001 (2006).

\bibitem{Treutlein06b}
P.~Treutlein {\itshape et al.}, Phys. Rev. A {\bfseries 74}, 022312 (2006).

\bibitem{Shore93}
B.~W.~Shore and P.~L.~Knight, J. Mod. Opt. {\bfseries 40}, 1195 (1993).

\bibitem{Kimble94}
H.~J.~Kimble, in \emph{Cavity Quantum Electrodynamics}, edited by P.~Berman (Academic Press, San Diego, 1994).

\bibitem{Katori04}
H.~Katori and T.~Akatsuka, Jpn. J. Appl. Phys. {\bfseries 43}, 358 (2004).


\end{thebibliography}
\end{document}